# Hadrons as Kerr-Newman Black Holes

Robert L. Oldershaw

Amherst College

Amherst, MA 01002

USA

rloldershaw@amherst.edu

20 Pages, 0 Figures, 1 Table

**Abstract:**

The scale invariance of the source-free Einstein field equations suggests that one might be able to model hadrons as "strong gravity" black holes, if one uses an appropriate rescaling of units or a revised gravitational coupling factor. The inner consistency of this hypothesis is tested by retrodicting a close approximation to the mass of the proton from an equation that relates the angular momentum and mass of a Kerr black hole. More accurate mass and radius values for the proton are then retrodicted using the geometrodynamic form of the full Kerr-Newman solution of the Einstein-Maxwell equations. The radius of an alpha particle is calculated as an additional retrodictive test. In a third retrodictive test of the "strong gravity" hypothesis, the subatomic particle mass spectrum in the 100 MeV to 7,000 MeV range is retrodicted to a first approximation using the Kerr solution of General Relativity. The particle masses appear to form a restricted set of quantized values of the Kerr solution: $n^{1/2} \mathfrak{M}$, where values of n are a set of discrete integers and $\mathfrak{M}$ is the revised Planck mass. The accuracy of the 27 retrodicted masses averages 98.4%. Finally, the new atomic scale gravitational coupling constant suggests a radical revision of the assumptions governing the Planck scale, and leads to a natural explanation for the fine structure constant.

**Key Words:** hadrons, black holes, Kerr solution, self-similarity, fractals, Planck scale, cosmology

# 1. Introduction

The Einstein field equations of General Relativity can be written (Misner, Thorne and Wheeler, 1973) as:

$$R_{\mu\nu} - 1/2\, g_{\mu\nu} R = k\, T_{\mu\nu} \quad (1)$$

where $R_{\mu\nu}$ is the Ricci tensor, $g_{\mu\nu}$ is the metric tensor, R is the Ricci scalar, $T_{\mu\nu}$ is the stress-energy tensor and k is the coupling factor between the geometry of a space-time and its matter content. This equation can be written in an even more compact form:

$$G_{\mu\nu} = k\, T_{\mu\nu} \quad (2)$$

where $G_{\mu\nu}$ is called the Einstein tensor. This deceptively simple expression disguises the fact that the equation represents a complicated and coupled set of 10 nonlinear partial differential equations in 4 unknowns. However, the conceptual meaning of the equation has a simple elegance. The geometry of the space-time ($G_{\mu\nu}$) is determined by the energy and momentum densities/fluxes of the matter ($T_{\mu\nu}$) and, reciprocally, the motions of the matter are determined by the geometry of the space-time.

For the purposes of the present discussion, we focus primarily on the term k in Eqs. (1) and (2). Einstein, and those who have followed in his footsteps, noted that if

$$k = (8\pi/c^4)\, G \quad (3)$$

where G is the conventional Newtonian gravitational constant and c is the velocity of light, then General Relativity successfully predicts the observed advance of the perihelion of Mercury, the deflection of light rays passing near to the Sun, the observed frame-dragging of space-time, and all the usual macroscopic results approximated by Newtonian gravitation. However, in the late 1970s and early 1980s several physicists including A. Salam and J. Strathdee (1978), C. Sivaram and K. P. Sinha (1977), E. Recami and P. Castorina (1976, 2005) and M. Pavsic (1978) explored the theoretical possibility of "strong gravity" within the microcosm.

## 2. "Strong Gravity"

The rationale for "strong gravity" is as follows. The vacuum equations of both General Relativity and Maxwell's equations are manifestly scale invariant. If masses, charges and dimensional "constants" are to be included in a discretely scale invariant version of General Relativity, then they would have to be suitably scaled. One can hypothesize a global discrete dilation invariance (a component of conformal symmetry) wherein all length (L), time (T) and mass (M) *units* scale according to *discrete* transformations of the form:

$$L_\Psi = \Lambda \, L_{\Psi-1} \quad (4)$$

$$T_\Psi = \Lambda \, T_{\Psi-1} \quad (5)$$

$$M_\Psi = \Lambda^D \, M_{\Psi-1} \quad (6)$$

where $\Lambda$ and $D$ are dimensionless scaling constants, and where $\Psi$ and $\Psi-1$ designate two neighboring discrete scales with a self-similar hierarchical arrangement, i.e., macrocosm ($\Psi = 0$) and microcosm ($\Psi = -1$). Going back to Eq. (2), we can consider the idea that, whereas k applies in the macrocosm,

$$k' = (8\pi/c^4)(\Lambda^{D-1} G_0) = (8\pi/c^4) \, G_{-1} \quad (7)$$

applies in the subatomic realm. The term $\Lambda^{D-1}$ arises because the dimensionality of $G_\Psi$ is $L^3/MT^2$, and therefore $G_{-1} = \Lambda^{D+2-3} G_0$. For the case of hadrons, it is at least logically possible that the gravitational coupling between matter and the geometry of space-time is much stronger than for macroscopic systems.  The value of the gravitational coupling factor has never been measured *within* an atom or a subatomic particle.  The standard use of the Newtonian value in this domain is based purely on an untested assumption.

One might well ask whether there are observational data or theoretical results that support the "strong gravity" hypothesis.  In fact there is some interesting evidence that is consistent with this unorthodox idea.  As discussed in detail by Sivaram and Sinha (1977), hadrons and Kerr-Newman black holes share an intriguing set of similarities.

1. Both hadrons and Kerr-Newman black holes are almost entirely characterized by just three parameters: mass, charge and angular momentum.
2. Both hadrons and Kerr-Newman black holes have magnetic dipole moments, but do <u>not</u> have electric dipole moments.
3. Typical hadrons and Kerr-Newman black holes have gyromagnetic ratios of $\approx 2$.
4. Hadrons and Kerr-Newman black holes have similar linear relationships between angular momentum and mass squared, i.e., $J \propto M^2$.
5. When Kerr-Newman black holes interact, their surface areas may increase but can never decrease, which is potentially analogous to the increase of cross-sections found in hadron collisions.

Given these curious similarities between the fundamental characteristics of hadrons and Kerr-Newman black holes, there appears to be sufficient motivation for considering the "strong gravity" approach to hadrons.

## 3. Two Initial Retrodictive Tests

3.1 <u>The Mass and Radius of the Proton</u>

What we would like are some *empirical* tests of the primarily theoretical arguments for "strong gravity", and fortunately three appropriate retrodictive tests have been devised and evaluated. These quantitative tests all require a determination of $G_{-1}$ from Eq. (7). After analyzing a large and varied sample of comparative data from subatomic and stellar scale systems (Oldershaw, 1987, 1989a,b, 2001, 2007), the best empirical fit is found to be

$$G_{-1} = (\Lambda^{D-1} G_0) = 2.18 \times 10^{31} \text{ cm}^3/\text{g sec}^2 \qquad (8)$$

where $\Lambda = 5.2 \times 10^{17}$ and $D = 3.174$. The values for the dimensionless constants $\Lambda$ and $D$ were determined *decades before* the present tests were considered, and were derived using a *different set of observational data* (Oldershaw, 1989a,b)

For a macroscopic Kerr black hole there is an angular momentum (J) versus mass (M) relationship (McClintock, Shafee, Narayan, Remillard, Davis and Li, 2006) of the form:

$$J = aG_0M^2/c \qquad (9)$$

where a is the dimensionless spin parameter of the black hole and c is the velocity of light. If the "strong gravity" hypothesis has merit, then we should be able to apply a rescaled version of Eq. (9) to a stable hadron and get reasonable empirical results. Taking the proton as the archetypal "strong gravity" analogue for a macroscopic Kerr-Newman black hole, we have

$$j_{proton} \approx \hbar \approx (1/2) \, G_{-1} \, m^2/c \qquad (10)$$

where $\hbar$ is Planck's constant divided by $2\pi$, m is the proton mass and the dimensionless spin parameter is ½. Eq. (10) can be rearranged and reduced to

$$m \approx (hc/\pi \, G_{-1})^{1/2} \, . \qquad (11)$$

When we evaluate Eq. (11) for m, we get

$$m \approx [(6.63 \times 10^{-27} \, \text{erg sec})(2.99 \times 10^{10} \, \text{cm/sec})/\pi(2.18 \times 10^{31} \, \text{cm}^3/\text{g sec}^2)]^{1/2}$$

$$\approx 1.70 \times 10^{-24} \, \text{g} \, .$$

Our estimate for m based on the "strong gravity" hypothesis is in agreement with the measured proton mass of $1.67 \times 10^{-24}$ g at the 98.3% level. For a more accurate retrodictive test, the following method can be employed. Given $G_{-1}$ as the correct gravitational coupling factor for geometrizing mass, charge and specific angular momentum, we may employ the geometrodynamic approach (Misner, Thorne and Wheeler, 1973) to the the full Kerr-Newman solution of the Einstein-Maxwell equations. Calculations (Oldershaw, 2001) based on this more rigorous method yield the following values for the radius and mass of the proton.

$$r = m + [m^2 - q^2 - a^2 \cos^2 \theta]^{1/2} = 8.13 \times 10^{-14} \, \text{cm} \qquad (12)$$

$$m = \{[m_{ir} + q^2/4m_{ir}]^2 + j^2/4(m_{ir})^2\}^{1/2} = 1.67 \times 10^{-24} \, \text{g} \qquad (13)$$

This demonstrates that the full Kerr-Newman solution of the Einstein-Maxwell equations accurately models the basic r, m, q and j properties of the proton when $G_{-1}$ is adopted as the correct gravitational coupling factor within atomic scale systems. For atomic nuclei with half-integer spin (a > 0), we have the approximate general relation

$$m_A \approx A(\hbar c/aG_{-1})^{1/2} \qquad (14)$$

where A is the atomic mass, i.e., the number of nucleons.

3.2  The Radius of the Alpha Particle

A second retrodictive test involves a radius estimate rather than a mass estimate. The alpha particle ($^4$He$^{++}$) has a spin of 0 and therefore we cannot use the Kerr-Newman solution of General Relativity for this test. However, we can employ a Schwarzschild solution for an order-of-magnitude check on the appropriateness of $G_{-1}$ in the case of a system that is more massive than the proton and has $a = 0$. For the Schwarzschild metric in the stellar scale context

$$R = 2G_0M/c^2 \qquad (15)$$

where R is the radius of a Schwarzschild black hole. In the case of an alpha particle we hypothesize that

$$r_\alpha \approx 2G_{-1}m_\alpha/c^2 \qquad (16)$$

where $r_\alpha$ and $m_\alpha$ are the radius and mass of the alpha particle, respectively. Evaluating Eq. (16), we get

$$r_{\alpha,Sch} \approx (2)(2.18 \times 10^{31} \text{ cm}^3/\text{g sec}^2)(6.68 \times 10^{-24} \text{ g})/(c)^2 \approx 3.26 \times 10^{-13} \text{ cm}.$$

The empirical value for $r_{\alpha,emp} \approx 2.2 \times 10^{-13}$ cm, and therefore $r_{\alpha,Sch} \approx 1.5 \, r_{\alpha,emp}$. Given that the Schwarzschild solution probably offers only a rough approximation for any actual physical system in nature, we have achieved a reasonable level of agreement between the observed radius of the alpha particle and an approximate theoretical estimate based on the "strong gravity" hypothesis.

## 4. Retrodicting the Subatomic Particle Mass Spectrum

4.1 Introduction

Retrodictions and predictions of subatomic particle masses have been highly valued *desiderata* ever since these unanticipated ultra-compact systems were discovered empirically. It is widely acknowledged that the particle masses have to be put into the Standard Model of particle physics "by hand", and further, that this lack of predictive/retrodictive capability is considered to be a significant problem that eventually must be resolved.

In this section we will consider the "strong gravity" approach to addressing the enigma posed by the particle mass spectrum. The main underlying idea of this approach is that gravitational interactions are stronger by a factor of ~ $10^{38}$ *within* subatomic particles than was previously realized. In subsection 4.3 the justification for this "strong gravity" approach will be developed in more detail. We consider the general hypothesis that these ultra-compact subatomic particles can be approximated as quantized allowed values, i.e., eigenstates or excited states, of the basic angular momentum-mass relationship of the Kerr solution of the Einstein field equations of General Relativity.

4.2 The Kerr Solution in the Subatomic Realm

We have seen above that the Kerr solution of the Einstein field equations of General Relativity yields the simplifying relationship $J = aGM^2/c$ between the angular momentum (J) of an ultra-compact object and its mass (M) (McClintock, Shafee, Narayan, Remillard, Davis, and Li, 2006). As in Eq. (9), the parameter a is referred to as a dimensionless spin parameter associated with the rotational properties of the ultra-compact object, G is the gravitational coupling factor and c is the velocity of light. Since we are interested in the masses of the ultra-compact objects, we rewrite Eq. (9) in the form:

$$M = (Jc/aG)^{1/2} . \qquad (17)$$

Since we want to apply Eq. (17) to the subatomic domain, we hypothesize that the unit of J in this domain is $\hbar$. Furthermore, in the subatomic realm we expect J to be restricted to a discrete set of values, i.e., $n\hbar$. As is well-known (Regge, 1959; Eden, 1971; Irving and Worden, 1977), in the 1960s Tullio Regge demonstrated that the masses and total spins of families of baryon and meson resonances were related by $J = kM^2$ relations. While Regge's heuristic phenomenology is well-documented, it has never found a fully adequate explanation in QCD or any other part of the Standard Model of particle physics.

With the above assumptions concerning J, we can rearrange Eq. (17) to yield:

$$M_n = n^{1/2} (\hbar c/G)^{1/2} , \qquad (18)$$

where n = 1/a, and we notice that $(\hbar c/G)^{1/2}$ is just the definition for the Planck mass. Therefore according to Eq. (18) the allowed values of the Kerr-derived J versus $M^2$ relation in the subatomic domain are the square roots of quantized multiples of the Planck mass. The applicability of the Kerr solution of General Relativity in the subatomic realm, and our initial assumptions concerning a and J, can be tested by attempting to retrodict the subatomic particle mass spectrum using Eq. (18).

## 4.3   Evaluating $(\hbar c/G)^{1/2}$ with Discrete Scale Relativity

The first step in testing Eq. (18) in the subatomic domain is to re-evaluate the Planck mass. Motivation for questioning the conventional Planck mass can be found in: (1) the fact that the conventional Planck mass (2.176 x $10^{-5}$ g) is not associated with any particle or phenomenon observed in nature, (2) the fact that the conventional Planck mass results in many forms of the closely related "hierarchy problem", and (3) the fact that the conventional Planck mass plays a major role in the vacuum energy density (VED) crisis in which there is a disparity of 120 orders of magnitude between the VED estimates of particle physics and cosmology.

A way to avoid these problems, and many more, can be found in a new cosmological paradigm for understanding nature's structural organization and dynamics (Oldershaw, 1989a,b, 2001). This new paradigm is called the Discrete Self-Similar Cosmological Paradigm (DSSCP). It is the product of a very thorough and careful empirical study of the physical properties of the actual objects that comprise nature, and the paradigm is based on the fundamental principle of discrete scale invariance. The discrete self-similar systems that comprise nature, and the fact that fractal structures are so common in nature, are the physical manifestations of the discrete scale invariance of nature's most fundamental laws and geometry. Discrete Scale Relativity (DSR) is the variation of the general DSSCP which postulates that the discrete self-similarity is *exact*, and has been discussed recently in a brief paper (Oldershaw, 2007). As we have seen in sections 2 and 3 above, DSR predicts that gravitation scales in the following discrete self-similar manner:

$$G_\Psi = (\Lambda^{1-D})^\Psi G_0 ,  \qquad (19)$$

where $G_0$ is the conventional Newtonian gravitational constant, $\Lambda$ and $D$ are empirically determined dimensionless self-similarity constants equaling $5.2 \times 10^{17}$ and $3.174$, respectively, and $\Psi$ is a discrete index denoting the specific cosmological scale under consideration. For the evaluation of Eq. (18) we have $\Psi = -1$, which designates the atomic scale, and therefore $G_{-1} = \Lambda^{2.174} G_0 = 2.18 \times 10^{31}$ cm$^3$/g sec$^2$. According to DSR, the proper gravitational coupling constant between matter and space-time geometry *within* atomic scale systems is $G_{-1}$. Evaluating the Planck mass relation $(\hbar c/G)^{1/2}$ using $G_{-1}$ and the usual values of $\hbar$ and c yields a value of $1.203 \times 10^{-24}$ g, or 674.8 MeV. This revised Planck mass is identified below by the symbol ₥. Therefore, if DSR and the "strong gravity" hypothesis are correct, then to a first approximation the subatomic particle mass spectrum should have peaks at the mass values:

$$M_n = n^{1/2} ₥ = n^{1/2} (674.8 \text{ MeV}) .  \qquad (20)$$

### 4.4  Testing $M_n = n^{1/2}$ ₥

Table 1 presents relevant data for testing Eq. (20) in terms of a *representative set* of subatomic particles from a mass/energy range of 100 MeV to 7,000 MeV. The particles appearing in Table 1 are among the most abundant, well-known and most stable members of the particle/resonance "zoo". For each integer of n there appears to be an associated particle, or set of related particles, that agrees with mass values generated by Eq. (20) at about the 93 to 99.99 % level. The *average* relative error for the full set of 27 particles is 1.6 %.

**Table 1**

**Representative Subatomic Particle Mass Spectrum (100 MeV to 7,000 MeV)**

| n | $n^{1/2}$ | $n^{1/2}$ (674.8 MeV) | Particle / MeV | Relative Error |
|---|---|---|---|---|
| 1/36 = (1/9)/4 | 0.1666 | 112.46 | μ / 105.66 | 6.4 % |
| 1/25 ≈ (1/6)/4 | 0.2000 | 134.96 | $\pi^0$ / 134.98 | 0.01 % |
| 1/2 = 2/4 | 0.7071 | 477.15 | κ / 497.65 | 4.1 % |
| 3/4 | 0.8660 | 584.39 | η / 547.75 | 6.7 % |
| 1 = 4/4 | 1.0000 | 674.8 | ℳ / 674.8 | --- |
| 5/4 | 1.1284 | 761.40 | ρ / ~ 770 | 1.1 % |
| 5/4 | 1.1284 | 761.40 | ω / ~ 783 | 2.8 % |
| 2 | 1.4142 | 954.31 | $p^+$ / 938.27 | 1.7 % |
| 2 | 1.4142 | 954.31 | n / 939.57 | 1.6 % |
| 2 | 1.4142 | 954.31 | η' / 957.75 | 0.4 % |
| 3 | 1.7320 | 1167.75 | $\Lambda^0$ / 1115.68 | 4.7 % |
| 3 | 1.7320 | 1167.75 | $\Sigma^i$ / <1192> | 2.0 % |
| 4 | 2.0000 | 1349.60 | $\Xi^0$ / 1314.83 | 2.6 % |
| 5 | 2.236 | 1508.90 | N(1440)/ 1430-1470 | ~ 4.8 % |
| 6 | 2.4495 | 1652.91 | $\Omega^-$ / 1672.45 | 1.2 % |
| 7 | 2.6458 | 1785.35 | $\tau^-$ / 1784.1 | 0.05 % |
| 8 | 2.8284 | 1908.62 | $D^0$ / 1864.5 | 2.4 % |
| 8 | 2.8284 | 1908.62 | $D^{+/-}$ /1869.3 | 2.1 % |
| 8 | 2.8284 | 1908.62 | $^2H$ / 1889.77 | 1.0 % |
| 10 | 3.1623 | 2133.90 | $D_s^i$ / 2112.1 | 1.0 % |
| 12 | 3.4641 | 2337.58 | $\Lambda_c^i$ / 2284.9 | 2.3 % |
| 14 | 3.7417 | 2524.87 | $\Xi_c^i$ / <2522.75> | ~ 0.1 % |
| 16 | 4.0000 | 2699.20 | $\Omega_c^0$ / 2697.5 | 0.1 % |
| 18 | 4.2426 | 2862.93 | $^3H$ / 2829.87 | 1.2 % |
| 18 | 4.2426 | 2862.93 | $^3He$ / 2829.84 | 1.2 % |
| 30 | 5.4772 | 3696.03 | $^4He$ / 3727.38 | 0.9 % |
| 64 | 8.000 | 5398.40 | $B_j^1$ / <5313.25> | ~ 1.6 % |
| 90 | 9.4868 | 6401.71 | $B_c^i$ / <6400> | ~ 0.1 % |

Table 1 lists the values of n, the retrodicted masses, the empirical masses, and the relative errors for 27 subatomic particles. Here we will discuss these 27 test particles in two separate groups: those particles that have masses $\geq m_p$, where $m_p$ is the proton mass, and those particles that have masses $< m_p$. For the former group we see that integer values of n generate good first approximation retrodictions for the particle masses with $m_p \leq m < 7,000$ MeV. The average accuracy is 98.4 % and there appears to be a preponderance of even values of n

For the much smaller group of particles with $m < m_p$, the set of n-values is not as simple and regular as it is for the $m \geq m_p$ group. The unit ℳ obviously has n = 1 but other members of this group have fractional values of n. The $\mu, \pi^0, \kappa, \eta$, ℳ, $\rho$ and $\omega$ particles can be *assigned* n = (1/9)/4, (1/6)/4, 2/4, 3/4, 4/4, 5/4 and 5/4, respectively, or n = 1/36, 1/25, 1/2, 3/4, 1, 5/4 and 5/4. One gets the definite impression that there is an underlying order to this set of n-values, but a unique pattern is not obvious. The distinct possibility exists that n-values for the m < $m_p$ group are *compound terms* such as n = i / j, or i · j, where i and j could be integers, multiples of $\pi$, and/or multiple rational fractions, e.g., $n_\eta$ = [3/2 · 1/2]. Rather than explore these possibilities numerologically, an approach with a long and checkered history, it seems more prudent to wait for a second approximation analysis of the subatomic particle mass spectrum using the *Kerr-Newman* solution of the Einstein-Maxwell equations to provide a more sophisticated model of the particles. This more complete and rigorous analysis would include charge, mass, electrodynamic considerations and spin-related phenomena. The results of this second approximation analysis should provide considerable guidance in understanding the most appropriate set(s) of n-values for all particles, as well as fostering an understanding the more subtle properties of the underlying order that generates the very regular patterns observed in the particle mass spectrum.

### 4.5 Implications of the Particle Mass Spectrum Retrodiction

(a) As demonstrated by the results presented in Table 1, the subatomic particle mass spectrum appears to manifest a simple, consistent and orderly pattern extending over a considerable range of particle masses and a diversity of family types, i.e., leptons, mesons, and baryons.

(b) To a first approximation the unit mass 𝔐 and the discrete angular momenta of the particles appear to be the primary or dominant physical determinants of the particle mass spectrum. Charge and other physical phenomena appear to be second order effects that determine the fine structure of the mass spectrum.

(c) A critical factor in determining the first approximation mass spectrum is the revised Planck mass ($\approx$ 674.8 MeV) which is uniquely obtained via the scaling relations of the Discrete Self-Similar Cosmological Paradigm.

(d) In atomic and nuclear physics, there are well-known examples (Rohlf, 1994; Garai, 2007) of phenomena wherein "magic numbers" appear in the stable solutions of the fundamental equations. This is especially evident in the isotopic stability of subatomic nuclei, and in the filling of electron "shells" in atoms. Perhaps the results shown in Table 1 identify an analogous case of a "magic numbers" phenomenon that applies in the context of the subatomic particle mass spectrum.

(e) A second approximation of the particle mass spectrum will clearly require the Kerr-Newman solution of the Einstein-Maxwell equations in order to fully take charge, spin and related phenomena into account. It can be *predicted* on the basis of the results discussed in this paper that the full Kerr-Newman solution will yield a more accurate retrodiction of the mass spectrum that includes more of the spectrum's fine structure. The geometrodynamics approach to working with the Kerr-Newman solution, as developed by Misner, Thorne and Wheeler (1973), would seem to offer a simple method for conducting initial tests of this prediction. Interested readers are strongly encouraged to participate in this effort.

## 5. Conclusions

Given the fact that $G_{-1}$ is evaluated empirically and is thus an approximation, the general agreement between the empirical and theoretical values for the proton mass, the proton radius, the alpha particle radius, and the subatomic particle mass spectrum encourages one to think that the hypothesis of "strong gravity", and the global discrete self-similarity of Discrete Scale Relativity, are worth pursuing. A follow-up (Oldershaw, 2010) to the present research demonstrates that the proposed revision of the k term in General Relativity leads to a radical reconsideration of assumptions involved in determining the Planck Scale, which is the microscopic scale at which General Relativity and Quantum Mechanics play equally important dynamical roles. If $G_{-1}$ is used in place of $G_0$ when one calculates the Planck length, Planck mass and Planck time, then the results are as follows.

$$\text{Planck length} \equiv (\hbar G_{-1}/c^3)^{1/2} = 2.9 \times 10^{-14} \text{ cm} \approx 0.4 \, r_{proton}. \qquad (21)$$

$$\text{Planck mass} \equiv (\hbar c/G_{-1})^{1/2} = 1.2 \times 10^{-24} \text{ g} \approx 0.7 \, m_{proton}. \qquad (22)$$

$$\text{Planck time} \equiv (\hbar G_{-1}/c^5)^{1/2} = 9.8 \times 10^{-25} \text{ sec} \approx 0.4 \, r_{proton}/c. \qquad (23)$$

When $G_{-1}$ and the revised Planck mass are substituted into the conventional equation for the fine structure constant, its value equals the ratio of the square of the unit electromagnetic charge to the square of the unit gravitational "charge." Equivalently, the fine structure constant equals the strength of the unit electromagnetic interaction divided by the strength of the unit gravitational interaction, for bound atomic scale systems (Oldershaw, 2009). If the "strong gravity" approach to atomic scale dynamics is correct, then the 80-year-old enigma of the physical meaning of the fine structure constant may have been resolved at last.

**Acknowledgement:** I would like to thank Dr. Jonathan Thornburg for helpful suggestions regarding the technical presentation of the research presented in section 4.